\documentclass{osa-article}
\journal{osac}
\articletype{Research Article}

\usepackage{graphicx}  % needed for figures
\usepackage{dcolumn}   % needed for some tables
\usepackage{bm}        % for math
\usepackage{verbatim}   % for math
\usepackage{upgreek}   % for math
\usepackage{hyperref}
\usepackage{braket}
\usepackage{lineno}
\usepackage[load-configurations=abbreviations, redefine-symbols=true]{siunitx}
\usepackage{tikz}

\begin{document}

\title{Cavity-enhanced non-destructive detection of atoms for an optical lattice clock}

\author{Richard~Hobson,\textsuperscript{1,*,+} William~Bowden,\textsuperscript{1,+}
Alvise~Vianello,\textsuperscript{1,2}
Ian~R.~Hill,\textsuperscript{1}
and Patrick~Gill\textsuperscript{1,2,3}}
\address{\textsuperscript{1}National Physical Laboratory, Hampton Road, Teddington TW11 0LW, United Kingdom\\
\textsuperscript{2}Blackett Laboratory, Imperial College London, Prince Consort Road, London SW7 2AZ, United Kingdom\\
\textsuperscript{3}Clarendon Laboratory, Parks Road, Oxford OX1 3PU, United Kingdom}
\affil{\textsuperscript{+}These authors contributed equally to this work}
\email{\textsuperscript{*}richard.hobson@npl.co.uk}

%\author{Richard~Hobson}
%\affiliation{National Physical Laboratory, Hampton Road, Teddington TW11 0LW, United Kingdom}

%\author{William~Bowden}
%\affiliation{National Physical Laboratory, Hampton Road, Teddington TW11 0LW, United Kingdom}

%\author{Alvise~Vianello}
%\affiliation{National Physical Laboratory, Hampton Road, Teddington TW11 0LW, United Kingdom}
%\affiliation{Blackett Laboratory, Imperial College London, Prince Consort Road, London SW7 2AZ, United Kingdom}

%\author{Ian~R.~Hill}
%\affiliation{National Physical Laboratory, Hampton Road, Teddington TW11 0LW, United Kingdom}

%\author{Patrick~Gill}
%\affiliation{National Physical Laboratory, Hampton Road, Teddington TW11 0LW, United Kingdom}
%\affiliation{Blackett Laboratory, Imperial College London, Prince Consort Road, London SW7 2AZ, United Kingdom}
%\affiliation{Clarendon Laboratory, Parks Road, Oxford OX1 3PU, United Kingdom}

\date{\today}% It is always \today, today,

\begin{abstract}
We demonstrate a new method of cavity-enhanced non-destructive detection of atoms for a strontium optical lattice clock. The detection scheme is shown to be linear in atom number up to at least \num{2e4} atoms, to reject technical noise sources, to achieve signal to noise ratio close to the photon shot noise limit, to provide spatially uniform atom-cavity coupling, and to minimize inhomogeneous ac Stark shifts. These features enable detection of atoms with minimal perturbation to the atomic state, a critical step towards realizing an ultra-high-stability, quantum-enhanced optical lattice clock.%: it facilitates the reduction of dead time by repeated recycling of atoms, and opens the door for collective spin-squeezing protocols to reduce quantum projection noise.
\end{abstract}

%\maketitle
\section{Introduction}

A range of quantum sensors from atom interferometers \cite{Cronin2009} to atomic clocks \cite{Ludlow2015} rely on spectroscopy of coherent atomic spins. These quantum sensors are extraordinarily precise, finding use in accurate frequency standards \cite{McGrew2018} and accelerometers for inertial navigation \cite{Wu2017}, but they also underpin several pioneering investigations of new physics beyond the standard model \cite{Delva2017,Rosi2017,ACME2018,Sanner2019}. However, these sensors are typically operated using samples of $N$ statistically uncorrelated atoms, meaning that their stability is affected by shot noise or quantum projection noise at the standard quantum limit \cite{Itano1993}. If the atomic samples were instead prepared in a correlated, spin squeezed quantum state, then the precision of quantum sensors could potentially be enhanced by a substantial factor of up to $\sqrt{N}$ \cite{Wineland1992a}.%, from violations of Lorentz invariance \cite{Delva2017} and the Einstein equivalence principle \cite{Rosi2017} to probes for an electron electric dipole moment \cite{ACME2014}.inertial sensors \cite{Dickerson2013}

%Spin-squeezed quantum states have the potential to enhance the precision of a range of quantum sensors \cite{Wineland1992a}, from atom interferometers \cite{Kovachy2015} to atomic clocks \cite{McGrew2018}. However, the preparation of squeezed states remains a significant technical challenge. Proofs of principle have been realized in microwave atomic clocks \cite{Leroux2010, Hosten2016}, but the optical lattice clock.

%In this work we explore a method of cavity-enhanced non-destructive detection of atoms in an optical lattice clock, which has the potential to prepare spin-squeezed states via weak measurement of the atomic spin.
In this work we take a step towards spin squeezing in an optical lattice clock, demonstrating a new scheme for cavity-enhanced non-destructive detection of strontium atoms. Cavity-enhanced detection methods have previously been applied to microwave atomic clocks, in which weak cavity-based measurements were used to prepare spin-squeezed states \cite{Hosten2016,Cox2016a} or to implement an atom phase lock \cite{Kohlhaas2015}. However, efforts to employ such techniques on optical lattice clocks, a highly accurate and stable configuration of atomic clock \cite{McGrew2018,Ushijima2015}, have so far been much too destructive to realize correlated quantum states \cite{Lodewyck2009,Vallet2017}.

The method presented here has been carefully designed to detect atoms in the lattice clock with large signal to noise and small perturbation to the atomic state. The signal is found to be linear in atom number over a wide range, enabling the use of large samples $N \gtrsim \num{e4}$ where the potential stability gains from spin squeezing are greatest. Although technical noise sources in our implementation of the scheme have so far prevented reaching a signal to noise compatible with spin squeezing, we realize the first demonstration of recycling of atoms after detection. This is advantageous for the optical lattice clock since it can improve the duty cycle of clock interrogation, thus minimizing the Dick-effect frequency instability of the clock \cite{Dick1987,Westergaard2010}. In principle, taking the cavity-enhanced detection scheme to its fundamental photon shot noise limit should enable us to enter the quantum correlated regime in the optical lattice clock.%The technical noise sources which limit our implementation---spectral impurity of the probe laser and photodetector noise---are solvable with fairly minor hardware upgrades. If the scheme were upgraded to reach the photon shot noise limit, the quantum correlated regime would be within reach.

\section{Detection scheme}

The detection scheme presented here relies on the dispersion of light as it propagates through a sample of cold strontium trapped in a high-finesse cavity. The dispersion due to the $^1$S$_0$ to $^1$P$_1$ transition at \SI{461}{\nano\meter} causes the resonant frequencies of the cavity to shift in proportion to the number of atoms present in the cavity. By probing the atom-dressed cavity resonances as depicted in Figs. \ref{fig:Modulation_scheme} and \ref{fig:experimental_setup}, a signal can be synthesized which is proportional to the atom-induced cavity shift, and therefore also proportional to atom number. Here, we explain the operating principles of the detection scheme and we derive expressions to quantify the atom number signal, the photon shot noise, and the probe-induced Stark shift.

\subsection{The atom signal} \label{sec:atom_signal}

When a strontium atom in its ground state is placed inside the cavity, it interacts with each TEM$_{00}$ cavity mode near \SI{461}{\nano\meter} with a vacuum Rabi frequency of
\begin{equation}
2g_n = \sqrt{\frac{6 \lambda_n^2 c \Gamma}{\pi^2 w_0^2 L}}e^{-\frac{\rho^2}{w_0^2}}\cos\left(k_n z + \frac{n\pi}{2}\right)
\end{equation}
in angular frequency units, where $n$ is the longitudinal mode number, $\lambda_n = 2\pi/k_n$ is the wavelength of the cavity mode, $\Gamma = 2\pi\times\SI{30.2}{\mega\hertz}$ is the decay rate of the $^1$S$_0$ to $^1$P$_1$ atomic transition, $L$ and $w_0$ are the cavity length and mode waist respectively, and $z$ and $\rho$ are the axial and radial displacements of the atom from the center of the cavity mode \cite{Kimble1998}. The Gouy phase and the beam divergence have been neglected for simplicity.

If $N$ ground-state atoms are placed in the cavity, then the resonant frequency of each cavity mode will experience a perturbation
\begin{equation}
\delta \omega_{n} = \frac{N \left<{g_n^2}\right>}{\Delta_\mathrm{n}} \approx \frac{3 \lambda_n^2 c \Gamma}{4 \pi^2 w_0^2 L \Delta_\mathrm{n}} N
\end{equation}
where $\Delta_\mathrm{n} = \omega_\mathrm{n} - \omega_0$ is the angular frequency detuning of the empty cavity resonance from the atomic transition. For the approximation to hold, we assume a large cavity detuning $\Delta_\mathrm{n}^2 \gg N \left<g_n^2\right>$, tight radial confinement of atoms to a region $\rho \ll w_0$, and an effectively random atom position along $z$ so that $\left<\cos^2(k_nz)\right> = 1/2$. Experimentally, these constraints on $\rho$ and $z$ are set by an optical lattice trap at \SI{813}{\nano\meter} enhanced within the same cavity.

\begin{figure}
\begin{center}
\includegraphics[width=8.6cm]{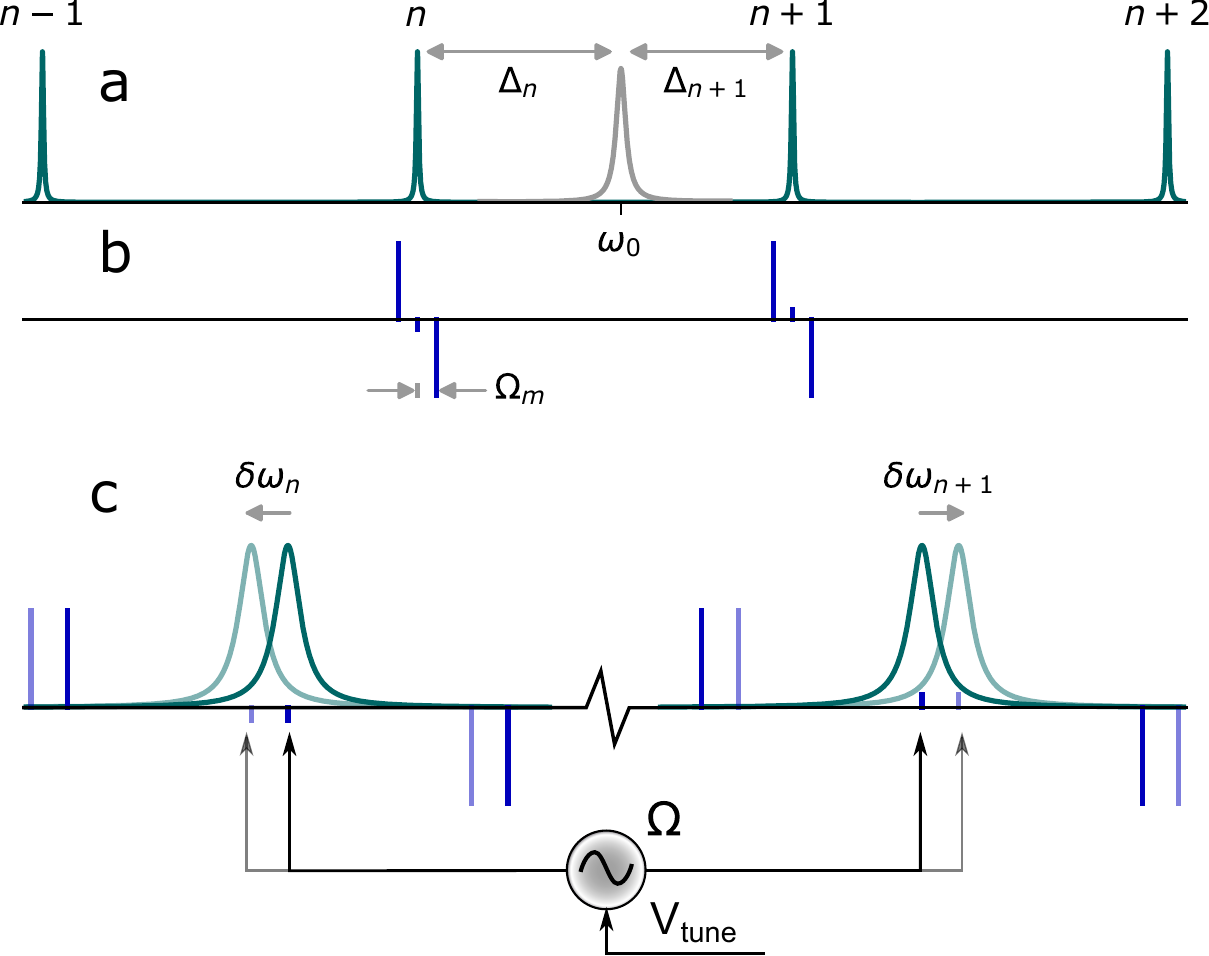}
\end{center}
\caption{(a) The cavity modes in green surround the atomic transition in grey. (b) The cavity is probed with a set of six discrete frequencies generated by a waveguide amplitude modulator. (c) The nearest cavity modes are shifted by $\delta\omega_\mathrm{n}$ and $\delta\omega_\mathrm{n+1}$ when atoms are in the cavity. The cavity shift is tracked in a feedback loop by sending a frequency tuning voltage to the RF modulation source, creating a signal $V_\mathrm{tune}$ proportional to the intracavity atom number (see text). Ideally the cavity modes at \SI{461}{\nano\meter} would be placed symmetrically around the transition so that $\delta\omega_\mathrm{n+1} \approx - \delta\omega_\mathrm{n}$. However, there is also the constraint that the \SI{813}{\nano\meter} lattice must be operated within a few MHz of the magic wavelength to realize an accurate atomic clock, and must be resonant with the same cavity. In our experimental system the closest magic-wavelength cavity mode within the piezo tuning range has a residual offset at \SI{461}{\nano\meter} of $(\omega_{n+1} + \omega_{n})/2-\omega_0 = 2\pi \times -\SI{173}{\mega\hertz}$.}
\label{fig:Modulation_scheme}
\end{figure}

If the cavity is probed using a laser with an electric field amplitude $E_{p_n}$ at a frequency $\omega_{p_n}$ close to the atom-dressed cavity resonance $\omega_\mathrm{n} + \delta\omega_\mathrm{n}$, then the reflected field amplitude $E_{r_n} \propto E_{p_n}$ will depend strongly on the residual frequency offset between the probe and the atom-dressed cavity. Assuming a high finesse cavity with mirror transmission and loss coefficients $t_1,l_1,t_2,l_2 \ll 1$, and keeping within the regime of small residual offset $\left(\omega_\mathrm{n} + \delta\omega_\mathrm{n}\right) - \omega_{p_n} \ll (t_1^2 + l_1^2+t_2^2+l_2^2)/2\pi\times \omega_\mathrm{FSR}$, we find
\begin{align}
E_{r_n} &\approx \left(Z_0 - 2\pi j\frac{\left(1 - Z_0\right)^2}{t_1^2}\frac{\omega_\mathrm{n} + \delta\omega_\mathrm{n} - \omega_{p_n}}{\omega_\mathrm{FSR}}\right)E_{p_n} \label{eq:E_r}
\end{align}
where $j$ is the imaginary unit, $\omega_\mathrm{FSR} = 2\pi \times c/2L$ is the free spectral range of the cavity and $Z_0 = (l_\mathrm{tot}^2/t_1^2 - 1)/(l_\mathrm{tot}^2/t_1^2 + 1)$ is the cavity impedance matching coefficient, with $l_\mathrm{tot}^2 = l_1^2+t_2^2+l_2^2+2b$ describing the total round-trip losses from all sources except for transmission $t^2_1$ through the cavity input mirror. For completeness we include in this expression the single-pass atomic power absorption $b = 3N \lambda^2 \Gamma^2/(4 \pi^2 w_0^2 \Delta_\mathrm{n}^2)$, although for our cavity parameters given in Table \ref{tab:experimental_parameters} the atomic absorption loss is much smaller than other cavity losses as long as $N \ll \num{4e5}$.

\begin{table}
\caption{Atomic cavity properties at \SI{461}{\nano\meter}\label{tab:experimental_parameters}}
%\begin{ruledtabular}
\begin{tabular}{lccr}
Parameter & Symbol & Value & Units \\
\hline
Input mirror transmission & $t_1^2$ & 345 & ppm \\
Output mirror transmission & $t_2^2$ & 160 & ppm\\
Mirror absorption/scatter loss & $l_1^2 + l_2^2$ & $< 50$ & ppm \\
Impedance matching coefficient & $Z_0$ & -0.33 & - \\
Mode waist & $w_0$ & 75 & \si{\micro\meter}\\
Cavity length & $L$ & 35.9 & mm\\
Cavity resonance linewidth & $\kappa/2\pi$ & 330 & kHz\\
Free spectral range & $\omega_\mathrm{FSR}/2\pi$ & 4.18 & GHz\\
Modulation frequency & $\Omega_m/2\pi$ & $50.3$ & \si{\mega\hertz}\\
Vacuum Rabi frequency & $\sqrt{\left<g_n^2\right>}/2\pi$ & $340$ & \si{\kilo\hertz}\\
Atom-induced cavity shift & $\delta\Omega/2\pi$ & $55\times N$ & \si{\hertz}\\
Single-atom cooperativity & C & 0.045 & - \\
Critical atom number & $N_\mathrm{crit}$ & 28 & - \\
Local oscillator power & $P_\mathrm{LO}$ & $30$ & \si{\micro\watt}\\
Detector quantum efficiency & $\eta_\mathrm{det}$ & 0.92 & - \\
Propagation loss & $\eta_\mathrm{prop}$ & 0.72 & - \\
Coupling efficiency & $\xi$ & 0.9 & - \\
\end{tabular}
%\end{ruledtabular}
\end{table}

\begin{figure}
\begin{center}
\includegraphics[width=8.6cm]{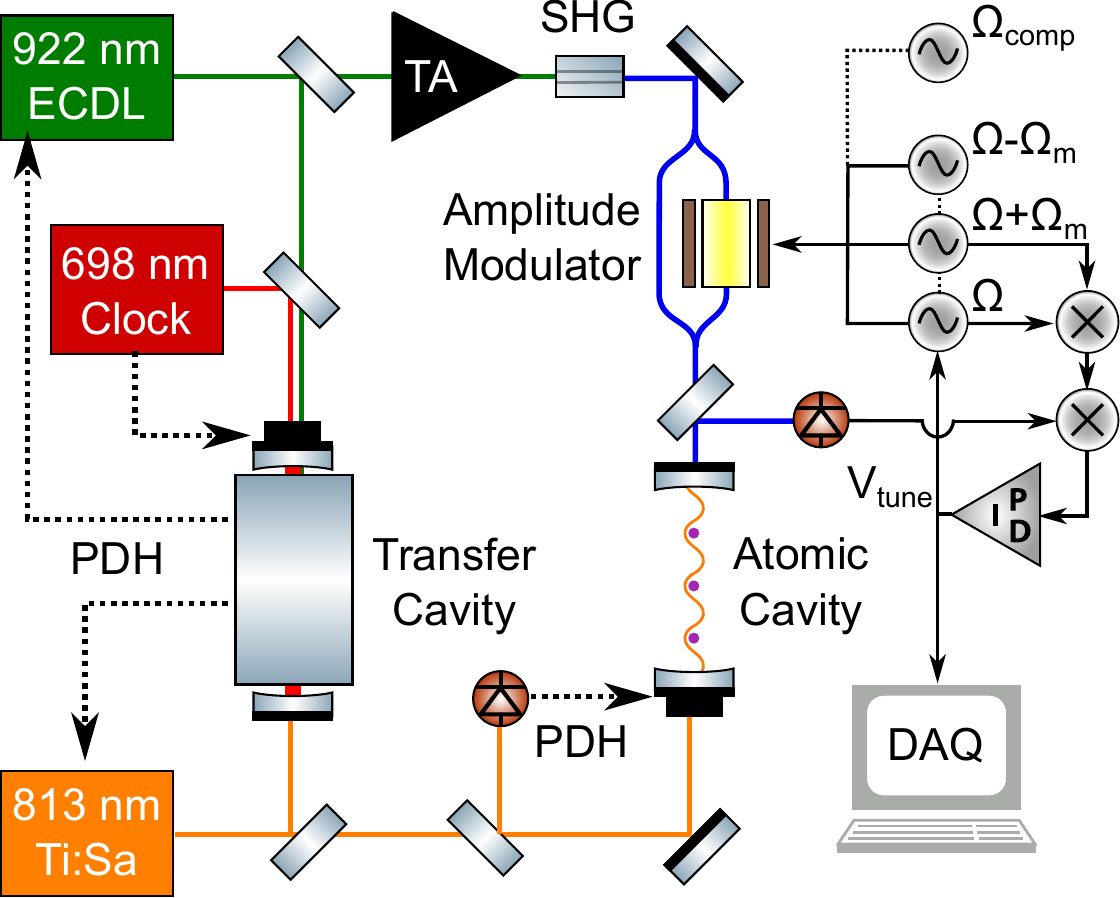}
\end{center}
\caption{The transfer cavity is used to lock the \SI{813}{\nano\meter} lattice and the \SI{461}{\nano\meter} probe to the \SI{698}{\nano\meter} clock laser frequency, ensuring that the \SI{461}{\nano\meter} probe is passively on resonance with the atomic cavity when the atomic cavity length is locked to the \SI{813}{\nano\meter} lattice. Definitions: TA: tapered amplifier, SHG: second harmonic generation, DAQ: data acquisition hardware (low noise oscilloscope), PDH: Pound-Drever-Hall frequency lock \cite{Drever1983}, $\Omega_\mathrm{comp}$: Modulation frequency for compensating inhomogeneous Stark shifts (see section \ref{sec:destructivity_and_shot_noise}), $\mathrm{V}_\mathrm{tune}$: frequency tuning voltage sent to the ``master'' oscillator $\Omega$, to which the other oscillators are phase locked with an offset of $\pm\Omega_m$.}
\label{fig:experimental_setup}
\end{figure}

To realize the non-destructive detection scheme explored in this work, two neighboring cavity modes $(n,n+1)$ are both probed simultaneously using the frequency spectrum depicted in Fig. \ref{fig:Modulation_scheme}. The six separate frequency components are generated using the optical setup in Fig. \ref{fig:experimental_setup}, in which a single \SI{461}{\nano\meter} beam at carrier frequency $\omega_c = (\omega_n + \omega_{n+1})/2$ is sent through an amplitude modulator to the cavity. The modulator is DC biased to generate zero carrier output power, and is then driven with the sum of three phase-locked voltage-controlled oscillators (VCOs) to modulate the light as:
\begin{equation}\label{eq:phase_modulation}
    \Phi(t) = A\left[\cos(\Omega + \Omega_m)t + \cos(\Omega - \Omega_m)t\right] + B\sin\Omega t
\end{equation}
for real amplitudes $B \ll A \ll 1$. The result is that very little of the optical power is present in the two cavity-coupled `probe' sidebands at $\omega_c \pm \Omega$, while most of the optical power exiting from the amplitude modulator is contained in the four 1\textsuperscript{st} order `local oscillator' sidebands at $\omega_c \pm \Omega \pm \Omega_m$. Further details of the amplitude modulator optics and drive electronics are given in appendix \ref{sec:appendix}. Note that it is important to suppress the transmission of carrier power through the amplitude modulator: even though it is far off resonance from the cavity modes, a poorly suppressed carrier could still leak into the cavity and cause significant excess scattering of photons due to its proximity to atomic resonance.

The purpose of probing two cavity modes $(n,n+1)$ simultaneously is to generate a beat signal at $\Omega_m$ whose amplitude is proportional to the frequency \textit{difference} between the two cavity resonances. This is achieved by constructive interference between two separate Pound-Drever-Hall (PDH) signals \cite{Drever1983}. The key advantage of this approach is that it provides first order immunity to technical noise on the laser frequency and the cavity length \cite{Ye1998,Long2007,Vallet2017}. This noise suppression means that stabilization of the probe laser to $\omega_c \approx (\omega_n + \omega_{n+1})/2$ at the few tens of \si{\kilo\hertz} level is sufficient to resolve differential atom-induced cavity shifts of a few hundreds of \si{\hertz}.

By analogy with the PDH scheme, the relative phases of the local oscillator and probe components of Eq. \eqref{eq:phase_modulation} are chosen so that, for each group of three frequency components, the beat signals at $\Omega_m$ of the two local oscillator components against $\Im(E_{r_n})$ or $\Im(E_{r_{n+1}})$ constructively interfere, while the less useful beats against $\Re(E_{r_n})$ or $\Re(E_{r_{n+1}})$ cancel out. Compared against the simpler approach where only one local oscillator modulation frequency is applied at either $\Omega + \Omega_m$ or $\Omega - \Omega_m$, this results in an extra factor of $\sqrt{2}$ in signal to photon shot noise ratio for the same total local oscillator power---an advantage also shared by some other related heterodyne detection schemes \cite{Beguin2014,Vallet2017}.

Propagating the cavity-coupled probe frequency components at $\omega_c \pm \Omega$ according to Eq \eqref{eq:E_r}, and assuming that the off-resonant local oscillator components at $\omega_c \pm \Omega \pm \Omega_m$ are reflected directly from the cavity input mirror (a valid assumption for large $\Omega_m$), a beat signal can be detected on a photodiode with an rms photocurrent amplitude $i(t) = i\sqrt{2}\cos(\Omega_m t)$:%the rms amplitude of the beat signal component of photocurrent $i(t) = i\sqrt{2}\cos(\Omega_m t)$ on the reflection photodetector can be written:
\begin{equation}
i \approx \frac{\xi\eta q_e}{\hbar\omega_0} \frac{\left(1 - Z_0\right)^2}{t_1^2} \frac{4\pi}{\omega_\mathrm{FSR}}\sqrt{P_\mathrm{p} P_\mathrm{LO}}\left(\frac{\delta\omega_{n+1} - \delta\omega_n}{2} - \delta\Omega\right)\label{eq:photocurrent}
\end{equation}
where $\delta\Omega = \Omega - \omega_\mathrm{FSR}/2$ is the frequency deviation of the probe oscillator from the half free spectral range of the empty cavity, $P_\mathrm{LO}/4$ and $P_\mathrm{p}/2$ are respectively the powers in each individual local oscillator and probe frequency component, $\eta = \eta_\mathrm{det}\eta_\mathrm{prop}$ is the detection efficiency of photons reflected from the cavity including photodiode quantum efficiency and propagation losses, $\xi$ is the coupling efficiency into the $\mathrm{TEM_{00}}$ cavity mode, and $q_e$ is the electron charge.

A simple implementation of the detection scheme could use static modulation frequencies with $\delta\Omega = 0$ and then read in the photocurrent $i$ as the atom number signal. However, we include one additional step to synthesize a signal proportional to the intracavity atom number: the photocurrent in Eq. \eqref{eq:photocurrent} is demodulated at $\Omega_m$ and sent into a feedback loop actuating on the frequency $\Omega$ to servo the beat amplitude $i$ to zero. The synthesizer generating $\Omega$ is set to have a base output of $\omega_\mathrm{FSR}/2$, so that the synthesizer frequency tuning voltage $V_\mathrm{tune}$ becomes proportional to the deviation $\delta\Omega \approx (\delta\omega_{n+1} - \delta\omega_n)/2$ once the loop is closed. This closed-loop configuration has two important advantages: (1) the size of the atom number signal $V_\mathrm{tune}$ is insensitive to probe and local oscillator power, and (2) the closed-loop signal is linear over a much larger range of atom numbers $N$.

\begin{figure}
\begin{center}
\includegraphics[width=9.6cm]{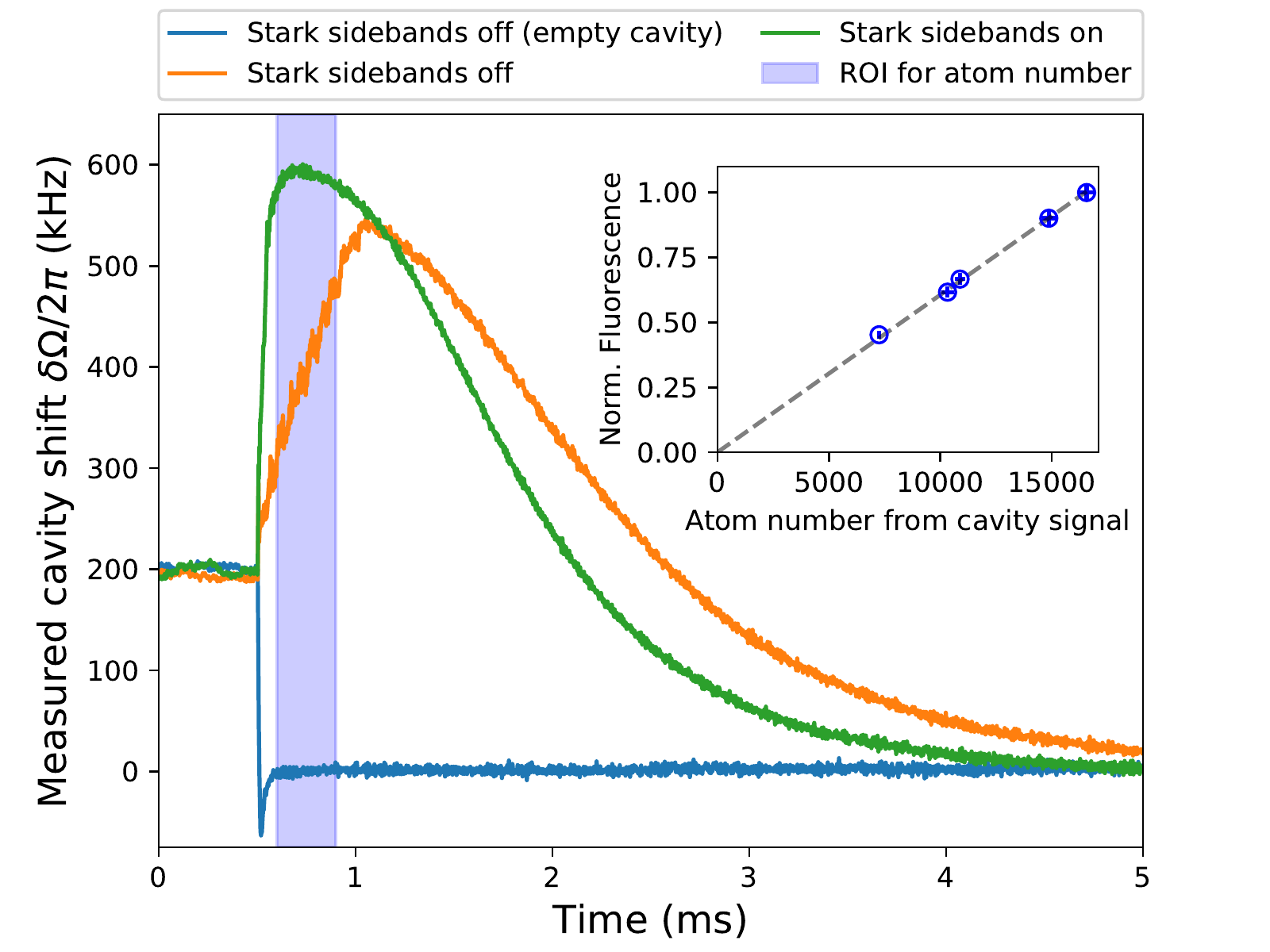}
\end{center}
\caption{Measurements of the intracavity atom number with and without additional Stark shift compensation sidebands being injected into the cavity on the mode pair $(n-3,n+4)$. The blue shaded region indicates the region of interest (ROI) used to count the atom number. The sidebands reduce the settling time of the lock (see Section \ref{sec:destructivity_and_shot_noise}), which is critical for being able to quickly measure the atom-induced cavity detuning non-destructively. Also shown is a background trace with no atoms and no compensation sidebands, verifying that the slow settling time is a result of atom dynamics induced by the inhomogeneous Stark shift. Inset: The linearity of the cavity-aided detection scheme is verified by comparing the atom-induced cavity detuning in the ROI against the fluorescence signal that is typically used to measure atom number.}
\label{fig:stark_comp}
\end{figure}

Experimental data based on $V_\mathrm{tune}$ is shown in Fig. \ref{fig:stark_comp}. The atom number $N$ is inferred by averaging $V_\mathrm{tune}$ in a region of interest lasting a few hundred microseconds. The signal is observed to be linear even for atom-induced cavity shifts much greater than the resonance linewidth of $\kappa/2\pi = \SI{330}{\kilo\hertz}$. To acquire each trace, a sample of up to \num{2e4} strontium atoms is prepared in the cavity-enhanced optical lattice trap at \SI{813}{\nano\meter}. The probe beam at \SI{461}{\nano\meter} is then switched on and $V_\mathrm{tune}$ is recorded. Before detection the atoms have been laser cooled in a two-stage magneto-optical trap \cite{Bowden2019}, reaching a final temperature of \SI{5}{\micro\kelvin} in the lattice. Since the atoms are cold compared with the lattice trap depth of \SI{15}{\micro\kelvin}, their radial spread is initially negligible compared with the waist of the non-destructive probe. However, once the probe is switched on, the atoms begin to heat up due to the scattering of \SI{461}{\nano\meter} photons. After a few milliseconds the atoms are hot enough to fly out of the probe region in the radial direction, resulting in the observed decay in the $V_\mathrm{tune}$ signal.

%Experimental data based on $V_\mathrm{tune}$ is shown in Figure \ref{fig:stark_comp}, demonstrating a linear signal even for atom-induced cavity shifts much greater than the resonance linewidth of \SI{330}{\kilo\hertz}. To acquire the traces, a sample of up to \num{2e4} strontium atoms is prepared in the cavity-enhanced optical lattice trap at \SI{813}{\nano\meter}, and then the probe beam at \SI{461}{\nano\meter} is switched on and $V_\mathrm{tune}$ is recorded. Before detection the atoms have been laser cooled using two stages of magneto-optical trap \cite{Hill2012}, reaching a final temperature of \SI{5}{\micro\kelvin} in the lattice. Since the atoms are cold compared with the lattice trap depth of \SI{15}{\micro\kelvin}, their radial spread is initially negligible compared with the waist of the non-destructive probe. However, once the probe is switched on, the atoms begin to heat up due to the scattering of \SI{461}{\nano\meter} photons. After a few \si{\milli\second} the atoms are hot enough to fly out of the probe region in the radial direction, resulting in the observed decay in the $V_\mathrm{tune}$ signal.

\subsection{Destructivity and shot noise} \label{sec:destructivity_and_shot_noise}

So far we have shown that the cavity-based detection scheme can generate a useful signal proportional to the atom number. However, in order to assess how non-destructive the scheme is, we need to characterize the effect of the probe light on the atomic state. Here, we consider two destructive processes: the scattering of probe photons and the probe ac Stark shift.

The probe and the local oscillator components all excite intracavity fields which interact with the atoms.
%, generating one-way circulating powers respectively given by $P^{(p)}_\mathrm{q} = P_\mathrm{p}/2 \times (1-Z_0)^2/t_1^2$ and $P^{(LO)}_q = P_\mathrm{LO}/2 \times t_1^2/4\pi^2 \times \omega_\mathrm{FSR}^2/\Omega_m^2$ in each of the two modes $q = n$ and $q = n+1$. However, as long as we use a sufficiently large modulation frequency $\Omega_m$ and avoid excessive local oscillator power, then the resonant probe components will dominate with $P^{(p)}_\mathrm{q} \gg P^{(LO)}_q$. The circulating probe power generates a standing-wave intensity profile for the two modes $q = n$ and $q = n+1$:
However, as long as we use moderate power in the local oscillators and a sufficiently large modulation frequency $\Omega_m$ compared with the cavity linewidth, then the resonant probe components will dominate inside the cavity. The intracavity circulating power is then given by $P^{(cav)}_\mathrm{q} = P_\mathrm{p}/2 \times (1-Z_0)^2/t_1^2$ in each of the two cavity modes $q = n$ and $q = n+1$, generating a standing-wave intensity profile in each mode:
\begin{align}
    I_q(\rho,z) &= \frac{8\xi}{\pi w_0^2}\frac{(1-Z_0)^2}{t_1^2}\frac{P_\mathrm{p}}{2}e^{-\frac{2\rho^2}{w_0^2}}\cos^2\left(k_q z + \frac{q\pi}{2}\right)
\end{align}
%for $q = n$ and $q = n+1$.

By applying steady state solutions to the optical Bloch equations \cite{SteckBook}, we find that the photon scattering rate per atom and the total ac Stark shift are given by
\begin{align}
    \Gamma_{sc}(\rho,z) &= \frac{3\lambda^3\Gamma^2}{16\pi^2\hbar c}\left(\frac{I_n(\rho,z)}{\Delta^2_n} + \frac{I_{n+1}(\rho,z)}{\Delta^2_{n+1}}\right) \\
    \Delta_{ac}(\rho,z) &= \frac{3\lambda^3\Gamma}{16\pi^2\hbar c}\left(\frac{I_n(\rho,z)}{\Delta_n} + \frac{I_{n+1}(\rho,z)}{\Delta_{n+1}}\right) \label{eq:ac_Stark}
\end{align}

Near $z = 0$ where the atoms are loaded, the neighboring cavity modes $(n,n+1)$ form two standing wave intensity profiles which are almost perfectly out of phase---one mode is shaped like $\sin^2(k_qz)$ while the other is $\cos^2(k_qz)$. As a result, for cavity modes arranged symmetrically around the atomic resonance such that $\Delta_n \approx -\Delta_{n+1}$, the total number of scattered probe photons is nearly independent of atom position along $z$ and can be approximated as
\begin{eqnarray}
\left<\Gamma_{sc} t\right> &\approx& \xi \left(\frac{1-Z_0}{t_1}\right)^2 \frac{3\lambda^2\Gamma^2}{2\pi^2 w_0^2 \Delta_{avg}^2} \frac{P_p t}{\hbar\omega_0} \label{eq:scatter} 
\end{eqnarray}
where $\Delta_{avg} = (\Delta_{n+1} - \Delta_{n})/2$ is the mean magnitude of the detuning from atomic resonance, $t$ is the probe time, and we have assumed tight radial confinement of atoms to a region $\rho \ll w_0$.

Equation \eqref{eq:scatter} highlights that a compromise must be made: in order to minimize the scattering of photons, we should avoid using large probe pulse areas $P_p t$. However, reducing the pulse area also reduces the signal size, and therefore compromises the signal to noise ratio of the atom number measurement. A fundamental signal to noise limit is imposed by photon shot noise from the intense local oscillator components, which causes a standard deviation in the integrated photocurrent $i \times t$ given by $\sigma_{i t} = \sqrt{\eta q_e P_\mathrm{LO} t/\hbar\omega_0}$. Comparing this shot noise against the signal discriminant in Eq. \eqref{eq:photocurrent}, and using Eq. \eqref{eq:scatter} to express $P_\mathrm{p}t$ as a function of scattered photons, we can derive the fundamental photon shot noise limit to the signal to noise ratio:
\begin{equation} \label{eq:SNR_phot}
\mathrm{SNR}_\mathrm{phot} = \frac{1-Z_0}{t_1}\frac{\lambda}{\pi w_0}N \sqrt{6\xi\eta\left<\Gamma_{sc} t\right>}
\end{equation}

In the context of optical lattice clocks, it is useful to compare the signal to photon shot noise ratio $\mathrm{SNR}_\mathrm{phot}$ against the signal to atom shot noise ratio $\mathrm{SNR}_\mathrm{SQL}$. The atom shot noise occurs when a projective measurement is applied to $N_\mathrm{TOT} \approx 2N$ atoms in an equal, coherent superposition of ground and metastable excited states. This type of shot noise imposes a ``standard quantum limit'' to the stability of the quantum sensor \cite{Wineland1992a}. For an ideal projective measurement, the measured ground-state atom number $N$ follows a binomial distribution with standard deviation $\sigma_N = \sqrt{N/2}$, resulting in a signal to atom shot noise ratio $\mathrm{SNR}_\mathrm{SQL} = \sqrt{2N}$. Combining this result with Eq. \eqref{eq:SNR_phot}, we observe that
\begin{equation}
    \frac{\mathrm{SNR}_\mathrm{phot}}{\mathrm{SNR}_\mathrm{SQL}} = \sqrt{\frac{N}{N_\mathrm{crit}}\left<\Gamma_{sc} t\right>}
\end{equation}
where the critical atom number $N_\mathrm{crit}$ is given by
\begin{equation}
     N_\mathrm{crit} = \left(\frac{t_1}{1-Z_0}\right)^2 \left(\frac{\pi w_0}{\lambda}\right)^2 \frac{1}{3\eta\xi}
\end{equation}
which represents the threshold number above which it is theoretically possible to perform weak measurement (scattering an average of less than one photon per atom) with noise below the standard quantum limit.

\begin{figure}
\begin{center}
\includegraphics[width=10cm]{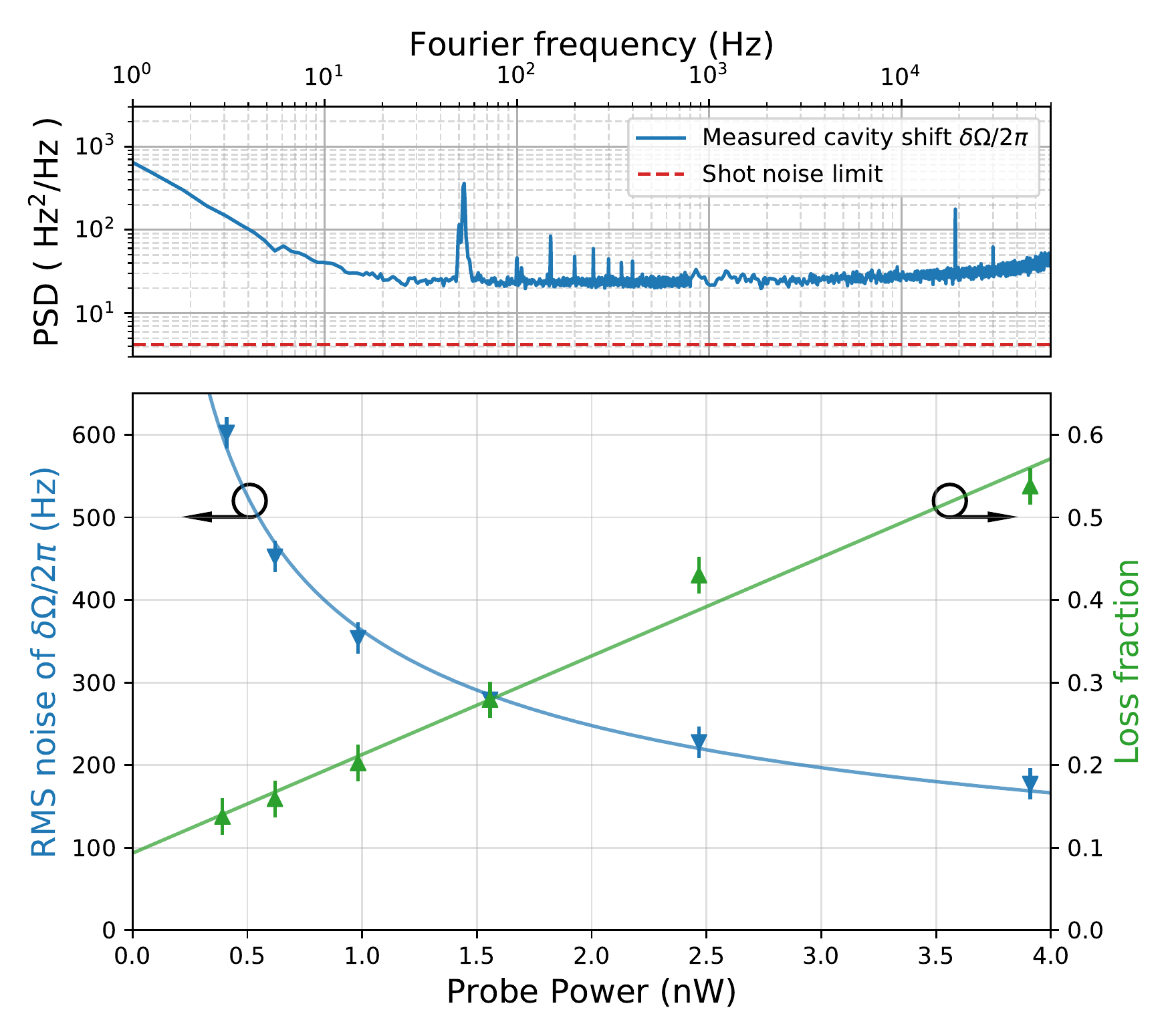}
\end{center}
\caption{\textit{Top:} Measured power spectral density (PSD) of noise on the atom-induced cavity shift $\delta\Omega/2\pi$ using probe power $P_p = \SI{1.6}{\nano\watt}$. The projected shot noise limit is also shown for reference. The noise peak at 50 Hz and subsequent harmonics result from a combination of electrical mains noise and a cooling fan close to the chamber, while the spike at \SI{19}{\kilo\hertz} corresponds to a dither frequency used to stabilize an etalon within the \SI{813}{\nano\meter} laser. \textit{Bottom:} Fraction of atoms lost from the lattice after a \SI{300}{\micro\second} probe pulse followed by a \SI{10}{\milli\second} delay, with $P_\mathrm{LO} = \SI{30}{\micro\watt}$ and a range of $P_p$. Also plotted is rms noise in $\delta\Omega/2\pi$ integrated over a bandwidth of \SI{10}{\hertz} to \SI{1.7}{\kilo\hertz}, with a fit curve proportional to $1/\sqrt{P_p}$.}
\label{fig:survival}
\end{figure}

For the experimental data presented in this work, the dominant noise sources were technical---primarily the imperfect photodetector and the amplified spontaneous emission from the tapered amplifier---which together introduce a white noise floor around six times higher than the photon shot noise limit. Nonetheless, as shown in Fig. \ref{fig:survival} the achieved signal to noise is quite sufficient to operate an atomic clock, even for probe powers weak enough to enable recycling of the atomic sample in the lattice.

In addition to the scattering of photons, the probe also induces an ac Stark shift $\Delta_{ac}(\rho,z)$ on the atomic ground state as given in Eq. \eqref{eq:ac_Stark}. Unlike the photon scatter rate, the ac Stark shift is not approximately uniform against position along the $z$ axis---the standing wave contributions to $\Delta_{ac}(\rho,z)$ have opposing signs, and therefore form a lattice dipole potential alternating between blue- and red-detuned intensity peaks. As observed in Fig. 3, at high probe power this lattice can have a substantial effect on the time-dependent atom number signal $V_\mathrm{tune}$---there are clear atomic motion settling dynamics during the first millisecond of the cavity interrogation. A speculative explanation of these settling dynamics is that the atoms are initially heated axially by the sudden presence of an intense standing-wave Stark shift, but then undergo axial Sisyphus cooling due to the interaction with the blue-detuned lattice \cite{Dalibard1985}, which slightly dominates over interactions with the red-detuned lattice since we operate with $|\Delta_{n+1}| < |\Delta_n|$. Regardless of their cause, the settling dynamics are inconvenient for atom number measurement: they reduce the signal size and add significant noise in the ROI.

In order to suppress the spatial dependence of the probe Stark shift along $z$, we apply an extra frequency component on the amplitude modulator at $\Omega_\mathrm{comp} = 7\omega_\mathrm{FSR}/4$ to generate second-order sidebands on the light at frequencies $\omega_c \pm 7\omega_\mathrm{FSR}/2$ \cite{Antypas2018, bennett1998}. The modulation amplitude at $\Omega_\mathrm{comp}$ is tuned to inject seven times as much power into the pair of cavity modes $(n-3,n+4)$ as is injected into modes $(n,n+1)$: this power ratio is chosen so that the standing-wave Stark shift from $(n-3,n+4)$ is the same size as the Stark shift from $(n,n+1)$, but has the opposite sign in the region near the center of the cavity $z \ll L/4$. An alternative approach could be to inject light into $(n-1, n+2)$, but we choose the further-detuned $(n-3,n+4)$ pair in order to scatter fewer photons for a given Stark shift and to reduce the interference between harmonics of $\Omega$ and $\Omega_\mathrm{comp}$. The effect of these additional Stark shift compensation sidebands is clear in Fig. \ref{fig:stark_comp}: when compensation sidebands are added the initial settling time is almost instant, limited by the bandwidth of the servo actuating on $V_\mathrm{tune}$ rather than by atomic motion dynamics. The improved settling time has the benefit of reducing the noise in the region of interest, thus ultimately improving signal to noise in the atom number readout. Additionally, suppression of the inhomogeneous Stark shift will prove to be essential in order to limit excess dephasing or `anti-squeezing' of the atomic state in future spin squeezing experiments \cite{Braverman2018}.

%We do not have a model or explanation for the \SI{461}{\nano\meter} lattice-induced dynamics---some Sisyphus cooling is to be expected in the $z$ direction because the atoms interact more strongly with the nearer detuned blue-detuned lattice in cavity mode $n+1$; however, it is unclear why an initially higher sample temperature would decrease the signal size.

\section{Conclusion}

We have developed and demonstrated a non-destructive detection scheme for optical lattice clocks. Compared with the similar cavity-based detection technique presented in Ref. \cite{Vallet2017}, we make several important advances toward realizing a quantum-enhanced clock: a short lock capture time, the recycling of atoms after non-destructive probe, compatibility with large atom number in the \num{e4} range, and compensation of inhomogeneous ac Stark shifts from the probe.

Future work will focus on two main applications. First we will exploit non-destructivity in the classical sense, minimizing the Dick instability of the optical lattice clock through repeated recycling of the atomic sample after detection \cite{Westergaard2010}. Second, we will take steps to reduce the detection noise toward the shot noise limit in order to realize minimally-perturbative non-destructive detection with less than one scattered photon per atom. In this regime of weak measurement it should be possible to prepare a spin-squeezed state with reduced quantum projection noise \cite{Wineland1992a}, replicating for optical clocks some of the metrological enhancements which have already been realized in microwave clocks using Rb \cite{Schleier-Smith2010a,Hosten2016, Cox2016a} and Cs \cite{Louchet-Chauvet2010}.

Ultimately, the relatively modest single-atom cooperativity $C = 4\left<g^2\right>/\kappa\Gamma = 0.045$ for the apparatus in this work will be a limiting factor for high detection efficiency at low atom number, as compared with previous demonstrations of high-fidelity cavity detection of Yb \cite{Braverman2019} and Sr \cite{Norcia2016a}. However, quantum non-destructive detection is still feasible in our cavity above the critical atom number $N_\mathrm{crit} = 28$, with significant metrological gain possible in the $N \approx \num{e4}$ range. Meanwhile, part of the motivation for lower cooperativity in this work is to include design features compatible with low and stable systematic shifts to the optical clock transition. Specifically, we choose cavity mirror coatings with high transmission at \SI{698}{\nano\meter} so that the clock transition can be excited uniformly along the lattice axis, minimizing residual $s$-wave collisions \cite{Blatt2009}; the lattice waist $w_0 = \SI{100}{\micro\meter}$ is selected to be large enough to trap high atom numbers with minimal density shifts; finally, the large \SI{18}{\milli\meter} separation of atoms from the cavity mirrors reduces vulnerability to patch charges, a known cause of dc Stark shifts \cite{Lodewyck2012}. The apparatus should therefore allow quantum-enhanced stability while still maintaining a systematic frequency uncertainty for the optical lattice clock in the \num{e-18} range \cite{Bowden2019}.

\section*{Funding}

This work was financially supported by the UK Department for Business, Energy and Industrial Strategy as part of the National Measurement System Programme; and by the European Metrology Programme for Innovation and Research (EMPIR) project 15SIB03-OC18. This project has received funding from the EMPIR programme co-financed by the Participating States and from the European Union's Horizon 2020 research and innovation programme.

\bibliography{library1}

\appendix
\section{Details of the amplitude modulator} \label{sec:appendix}

Here we explain some of the experimental details of the non-destructive detection scheme which were overlooked in Fig. \ref{fig:experimental_setup}. 

\begin{figure}
\begin{center}
\includegraphics[width=10cm]{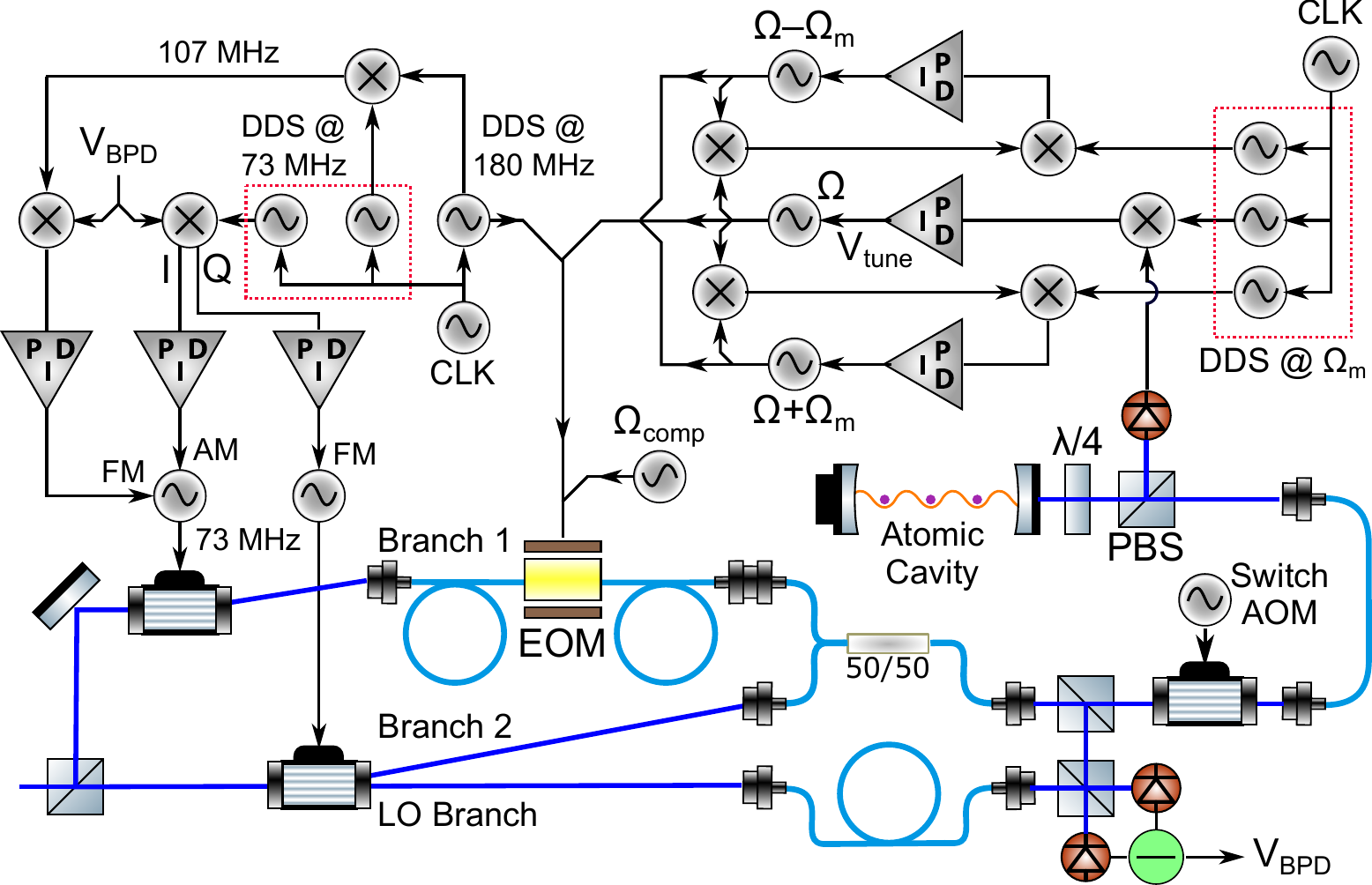}
\end{center}
\caption{\SI{461}{\nano\meter} optics and electronics underpinning the non-destructive detection scheme---see text for explanation. Several signal conditioning and beam steering components have been omitted from the diagram for the sake of readability, such as RF attenuators, filters and amplifiers, as well as mirrors, lenses and polarization control optics. Definitions: AOM: acousto-optic modulator, EOM: electro-optic modulator (commercial waveguide module in potassium titanyl phosphate), FM: frequency modulation, AM: amplitude modulation, DDS: direct digital synthesis, CLK: clock signal for DDS. \label{fig:QND_interferometer}}
\end{figure}

The non-destructive detection scheme centrally relies on a $\sim\SI{7}{\giga\hertz}$-bandwidth, high extinction ratio amplitude modulator, which is realized using the optics and electronics shown in Fig. \ref{fig:QND_interferometer}. We implement the amplitude modulator as a Mach-Zehnder interferometer with additional AOMs and EOMs to control the optical amplitude and phase of the individual branches. A number of feedback loops are employed to stabilize the overall phase delay of the modulator setup and to null the transmitted carrier.

The Mach-Zehnder interferometer is stabilized using the balanced photodiode signal $V_\mathrm{BPD}$ generated at the bottom right of the figure: weak \SI{180}{\mega\hertz} sidebands are injected onto the EOM in Branch 1 to create a beat signal at \SI{107}{\mega\hertz} on $V_\mathrm{BPD}$, which is then used to lock the optical phase of Branch 1 at a \SI{73}{\mega\hertz} offset from the LO branch. An I/Q demodulator forms the basis of two further control loops which are used to suppress both the I and the Q components of the strong beat remaining at \SI{73}{\mega\hertz}: the first loop acts on the phase of Branch 2 to ensure that the carriers exiting branches 1 and 2 have opposite phase; the second loop acts on the AOM diffraction efficiency of Branch 1 to match the carrier power in Branch 1 to the power in Branch 2. With the \SI{73}{\mega\hertz} beat suppressed, the interferometer transmits zero output power at the carrier frequency $\omega_c$, meaning that light sent to the atomic cavity is formed only of RF sidebands at $\omega_c \pm \Omega_\mathrm{mod}$ for the various RF modulation frequencies $\Omega_\mathrm{mod} \in \{\SI{180}{\mega\hertz},\Omega-\Omega_m,\Omega, \Omega+\Omega_m, \Omega_\mathrm{comp}\}$. The physical path lengths of Branch 1 and Branch 2 have been matched to within approximately \SI{1}{\milli\meter} to optimize the suppression of broadband laser noise exiting the interferometer.

The electronics underpinning the generation of the atom signal $V_\mathrm{tune}$ are depicted in the top right part of the figure. The two voltage-controlled oscillators (VCOs) at $\Omega \pm \Omega_m$ are offset-locked to a master oscillator at $\Omega$ with a loop bandwidth of \SI{1}{\mega\hertz} each. The offset locks are implemented using two separate DDS channels at $\Omega_m$, with a common clock and with relative phase tuned to match the optical phase modulation profile of Eq. \eqref{eq:phase_modulation}. In order to lock the master frequency $\Omega$ to the cavity mode splitting, the cavity reflection photodiode signal is demodulated at $\Omega_m$ and sent into a control loop acting on $\Omega$ with a unity gain bandwidth of \SI{150}{\kilo\hertz}. This master loop is always engaged, even when the \SI{461}{\nano\meter} probe beam is switched off, so to avoid railing the integrator there is a high-value DC limiting resistor in the feedback path. When the probe beam is switched on using the Switch AOM, the loop captures with a settling time of around \SI{20}{\micro\second} (see Fig. \ref{fig:stark_comp}) and the in-loop control voltage $V_\mathrm{tune}$ is finally read into a low noise PC oscilloscope to generate an estimate of atom number in each shot of the experiment.

%\begin{figure*}
%\begin{center}
%\includegraphics[width=13cm]{psd.pdf}
%\end{center}
%\caption{Power spectral density of the measured cavity shift using a \SI{1.6}{\micro\watt} of probe power. The spectrum exhibits a white noise floor of \SI{25}{\hertz^2\per\hertz} with flicker corner at \SI{10}{\hertz}. The spectral feature centred at 50 Hz and subsequent harmonics result from a combination of electrical mains noise and a cooling fan close to the chamber. The feature centred at \SI{19}{\kilo\hertz} is caused by the etalon modulation used to stabilize the Ti:Sapph resonator. \label{fig:QND_psd}}
%\end{figure*}
\end{document}